\documentclass[conference]{IEEEtran}
\IEEEoverridecommandlockouts

\usepackage{thmbox} 
\newtheorem[M, bodystyle=\normalfont\noindent]{definition}{Definition}

\newtheorem[M, bodystyle=\normalfont\noindent]{problem}{Problem}

\usepackage[ruled,vlined,shortend, linesnumbered]{algorithm2e} 
\usepackage[]{algorithmic} 
\usepackage{amssymb}

\usepackage{cite}
\usepackage{amsmath,amssymb,amsfonts}
\usepackage{algorithmic}
\usepackage{graphicx}
\usepackage{textcomp}
\usepackage{xcolor}
\usepackage[utf8]{inputenc}
\usepackage{newunicodechar}
\newunicodechar{−}{\textminus}
\usepackage{bm}
\usepackage{tabularx} 

\usepackage{mdwlist}
\usepackage[ruled,vlined,shortend, linesnumbered]{algorithm2e} 
\usepackage[]{algorithmic} 
\usepackage{booktabs}
\usepackage{color}

\usepackage{amsthm,amssymb}
\usepackage{amsmath}
\usepackage[font=small]{caption}
\usepackage[skip=3pt]{subcaption}
\usepackage{multirow}
\usepackage{mathrsfs}
\usepackage{amsbsy}
\usepackage[mathscr]{eucal} 
\usepackage[flushleft]{threeparttable}
\usepackage[hidelinks]{hyperref}
\usepackage{comment}
\usepackage{paralist}
\usepackage{pifont}

\usepackage{bbding} 

\usepackage{wrapfig}

\usepackage{enumitem}
\usepackage[protrusion=true,expansion=true]{microtype}
\pdfoutput=1

\def\BibTeX{{\rm B\kern-.05em{\sc i\kern-.025em b}\kern-.08em
    T\kern-.1667em\lower.7ex\hbox{E}\kern-.125emX}}

\newcommand{\hide}[1]{}

\newcommand{\bit}{\begin{compactitem}}
\newcommand{\eit}{\end{compactitem}}
\newcommand{\ben}{\begin{compactenum}}
\newcommand{\een}{\end{compactenum}}

\begin{document}

\title{ Frequency-Aware Sparse Optimization for Diagnosing Grid Instabilities and Collapses 
}

\author{Swadesh Vhakta$^1$,  Denis Osipov$^2$,  Reetam Sen Biswas$^3$, Amritanshu Pandey$^4$,\\ Seyyedali Hosseinalipour$^1$, and Shimiao Li$^1$\\
\thanks{Authors are with $^1$University at Buffalo--SUNY; $^2$New York Power Authority; $^3$GE Vernova Advanced Research; $^4$University of Vermont.
}
\\
\thanks{Manuscript accepted by PESGM 2026

©2026 IEEE. Personal use of this material is permitted. Permission from IEEE must be obtained for all other uses, in any current or future media, including reprinting/republishing this material for advertising or promotional purposes, creating new collective works, for resale or redistribution to servers or lists, or reuse of any copyrighted component of this work in other works.}
\vspace{-16mm}
}

\maketitle

\begin{abstract}
	This paper aims to proactively diagnose and manage frequency instability risks from a steady-state perspective, without running the complicated transient simulations. 
Specifically, we jointly address two questions \textit{(Q1) Survivability:} following a disturbance and the subsequent primary frequency response, can the system settle into a healthy steady state (feasible with an acceptable frequency deviation $\Delta f$)?  \textit{(Q2) Dominant Vulnerability:} if found unstable, what critical vulnerabilities create instability and/or full collapse? To address these questions, we first augment steady-state power flow states to include frequency-dependent governor relationships (i.e., governor power flow).
Afterwards, we propose a frequency-aware sparse optimization that finds the minimal set of bus locations with measurable compensations (corrective actions) to enforce power balance and maintain frequency within predefined/acceptable bounds. We evaluate our method on standard transmission systems to empirically validate its ability to localize dominant sources of vulnerabilities.
For a 1354-bus large system, our method detects compensations to only four buses under N-1 generation outage (3424.8 MW) while enforcing a maximum allowable steady-state frequency drop of 0.06 Hz (otherwise, frequency drops by nearly 0.08 Hz). 
We further validate the scalability of our method, requiring less than four minutes to obtain sparse solutions for the 1354-bus system.

\end{abstract}


\section{Introduction}
\label{sec:Introduction}
Understanding a power grid’s immediate survivability after disturbances (e.g., equipment failures or sudden load changes) is essential for ensuring system resilience. In line with this need, North American Electric Reliability Corporation (NERC) policies require N-1 resilience, i.e.,  systems must withstand and recover from N-1 contingency without major disruption. To ensure such resilience in practice, steady-state power-flow simulation \cite{powerflow-diverge} serves as the standard tool for identifying potential overloads and violations caused by equipment failures (e.g., generator trips). Building on this capability, contingency analysis is executed  within Energy Management Systems (EMS) during daily operations, at least every 30 minutes as required by NERC.
Beyond real-time monitoring, these simulations are also widely used in planning to evaluate future operating conditions and maintenance schedules. Nevertheless, with the increasing prevalence of extreme events (e.g., the 2021 Texas crisis \cite{2021TexasPowerCrisis}) and cyberattacks\cite{grid-security-Vyas} (e.g., MadIoT \cite{madiot,gridwarm}, Ukraine 2015-16 \cite{lee2017crashoverride}), there is an urgent need for diagnostic capabilities to provide deeper insights into grid vulnerabilities in the face of these emerging threats. 

To this end, understanding the limitations of traditional simulation tools becomes crucial. In this context, conventional steady-state simulators find a feasible operating point $\bm{v}^*$ by solving the nonlinear AC power-flow equations $\bm{g}(\bm{v})=0$. When a critical contingency (e.g., multiple generator trips) causes a large supply–demand imbalance, the equations may become infeasible, and the solver may diverge, signaling a system collapse. Yet, this approach has two key limitations.

\begin{wrapfigure}{r}{0.25\textwidth}
\vspace{-2mm}
\centering
\includegraphics[width=0.25\textwidth]{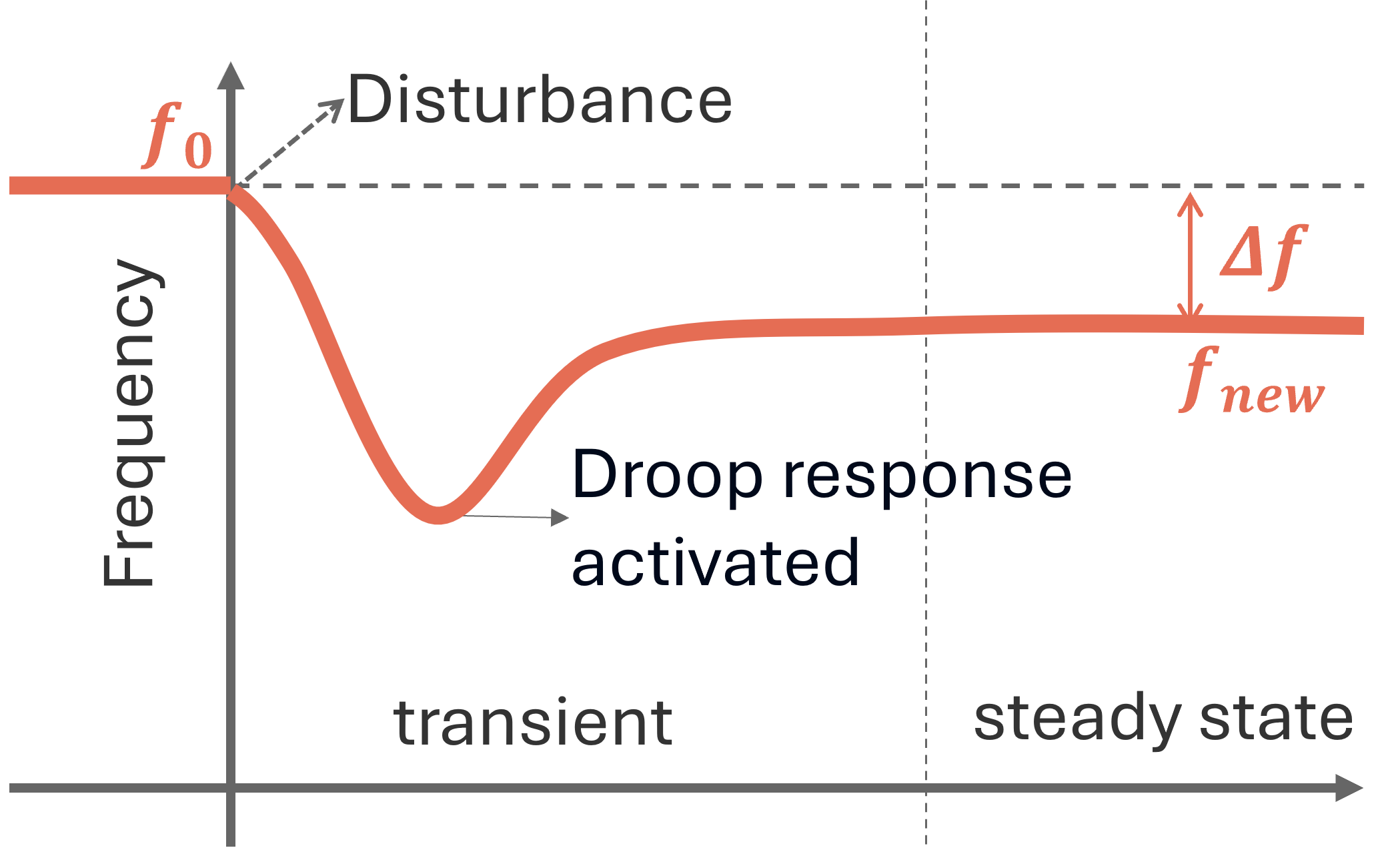}
\caption{Primary frequency response}
\label{fig:freq}
\vspace{-2mm}
\end{wrapfigure}
First, standard steady-state models do not capture the frequency state, which is essential for assessing instability risks. To illustrate this limitation, Fig. \ref{fig:freq} shows the realistic primary frequency response: immediately after a generation outage, the kinetic energy of rotating machines (and virtual inertia of inverters) temporarily supplies the power deficiency, causing a frequency drop; then governors act through droop control until a new steady-state is reached with frequency deviation $\Delta f$. Our focus is on this primary frequency response timescale (typically from a few seconds to one minute after disturbance), before slower secondary and tertiary (re-dispatch) controls take effect. While transient analysis equipped with generator governor models can capture such frequency response dynamics, it is computationally expensive {\cite{wu2023transient}. In some industry practices (e.g., CAISO \cite{caiso-CA}), the droop-response effect is approximated within a steady-state power flow study by redistributing the lost generation among droop-capable units using generation distribution factors, thereby accounting for droop response when simulating the system’s steady-state response to generator contingencies. Recent studies \cite{sugar-pf-freq} augmented steady-state analysis with implicit frequency modeling to estimate the final steady-state deviation $\Delta f$, and accurately assess post-droop stability.

\begin{figure*}[htbp]
    \centering
         \begin{subfigure}[h]{0.3\linewidth}
        \centering
        \includegraphics[width=0.95\linewidth]{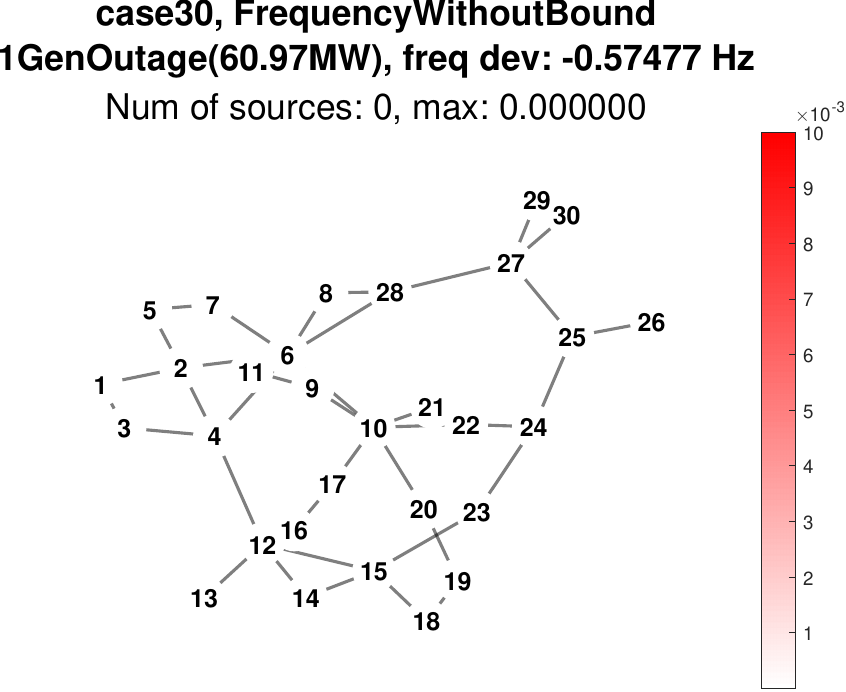}
        \caption{System is feasible after contingency.}
    \end{subfigure}
      \begin{subfigure}[h]{0.3\linewidth}
        \centering
        \includegraphics[width=0.95\linewidth]{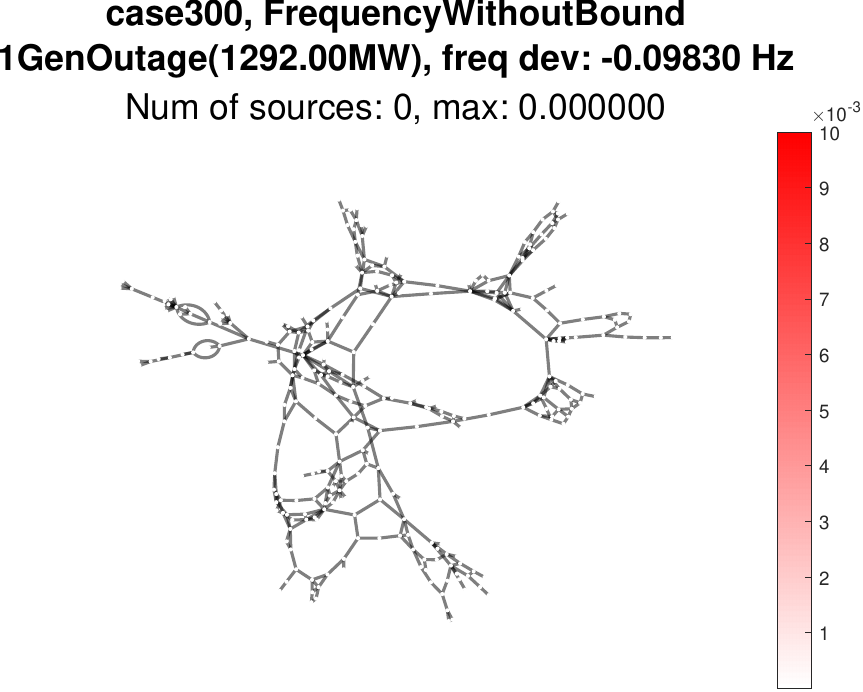}
        \caption{System is feasible after contingency.}
    \end{subfigure}
     \begin{subfigure}[h]{0.3\linewidth}
        \centering
        \includegraphics[width=0.95\linewidth]{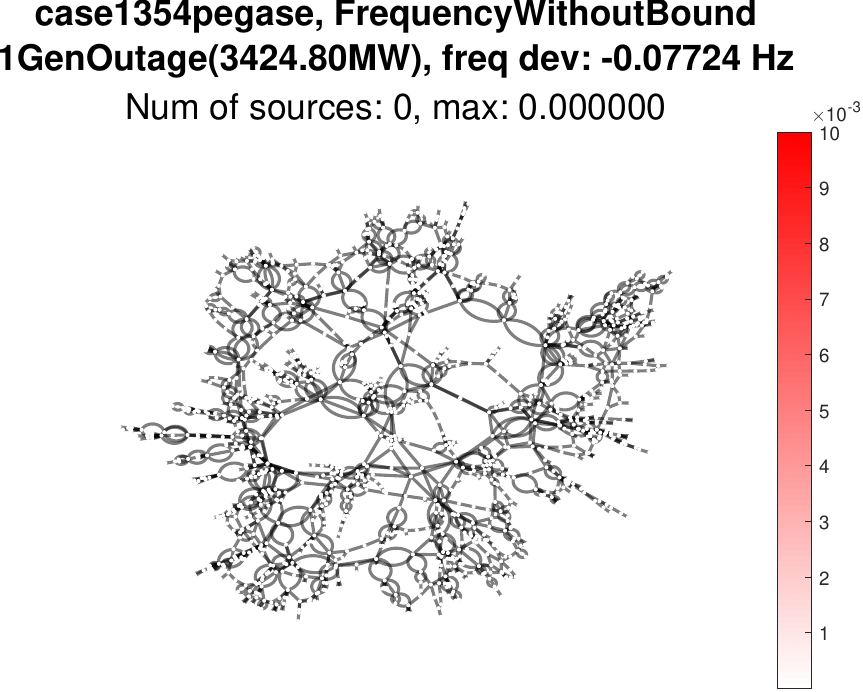}
        \caption{System is feasible after contingency.}
    \end{subfigure}    \hfill\\
    \begin{subfigure}[h]{0.3\linewidth}
        \centering
        \includegraphics[width=0.95\linewidth]{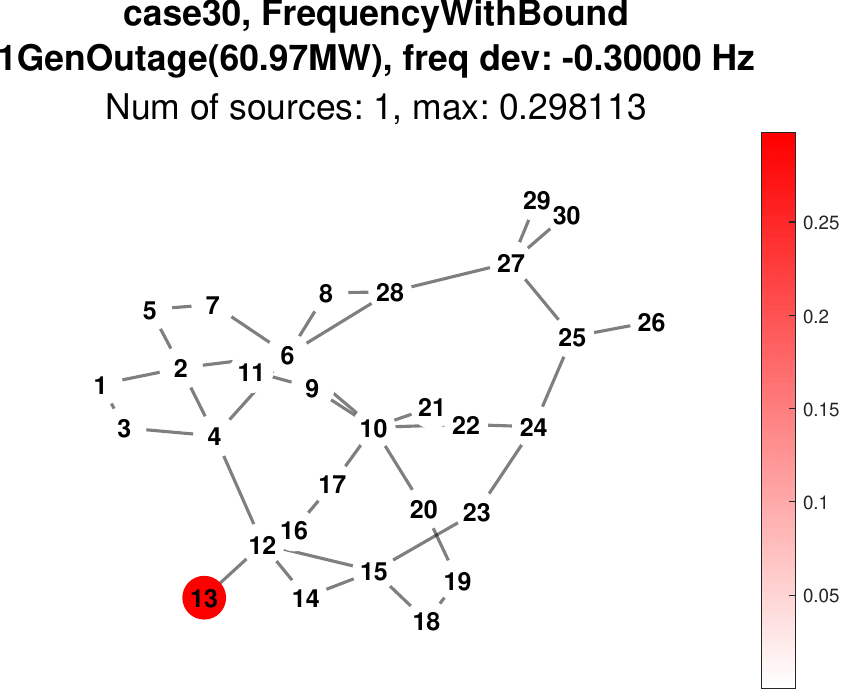}
        \caption{Enforce $-0.3Hz\leq\Delta f$}
        \label{2d}
    \end{subfigure}
    \begin{subfigure}[h]{0.3\linewidth}
        \centering
        \includegraphics[width=0.95\linewidth]{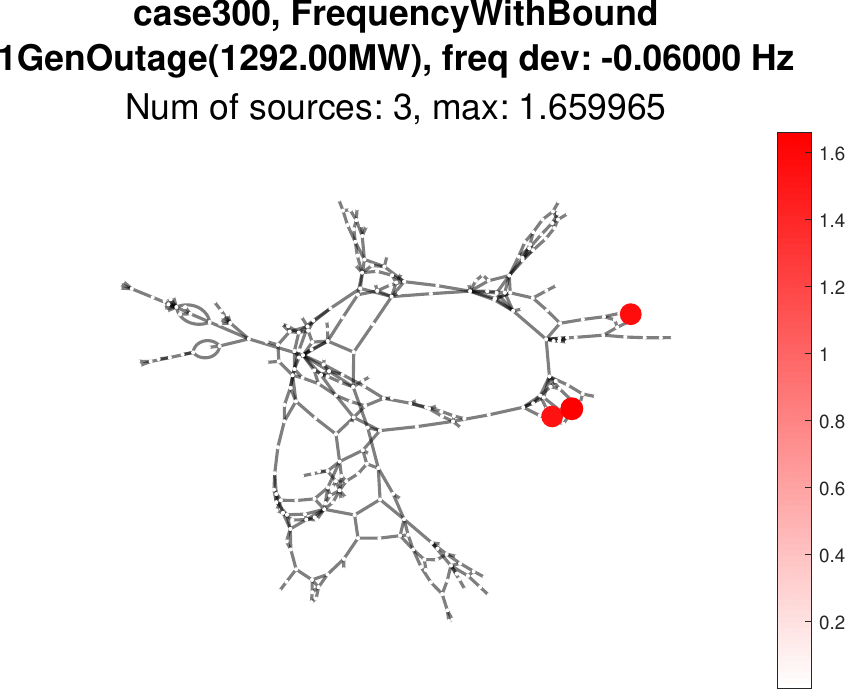}
        \caption{Enforce $-0.06Hz\leq\Delta f$}
        \label{2e}
    \end{subfigure}       
    \begin{subfigure}[h]{0.3\linewidth}
        \centering
        \includegraphics[width=0.95\linewidth]{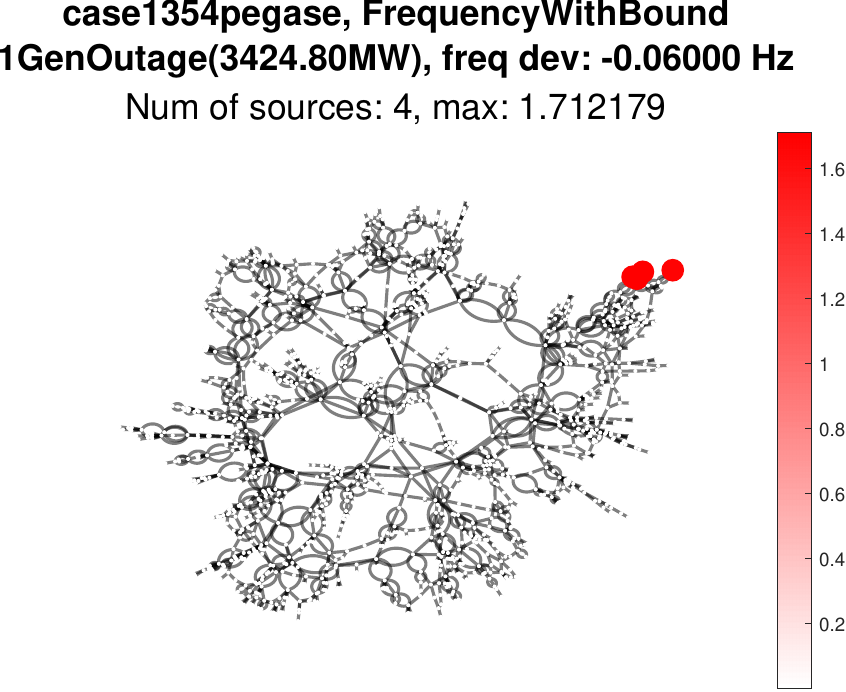}
        \caption{Enforce $-0.06Hz\leq\Delta f$}
        \label{2f}
    \end{subfigure}\hfill
    \caption{A few localized compensations can stabilize the entire system. Top row (baseline): simulation of steady-state frequency deviation after generation outage, with droop response considered. Bottom row (proposed method): Red nodes denote the magnitude of compensation sources (in per-unit current) needed at a few identified nodes to restore the $\Delta f$ within bound, localizing and quantifying instability vulnerabilities.}
    \label{fig: graph plot, sparsefeas-freq}
    \vspace{-4mm}
\end{figure*}

Second, under large instabilities and collapses, standard simulators that diverge provide little insight into the severity of collapse, the locations of \textit{dominant sources of vulnerabilities}, and the corrective actions required to restore feasibility. To address this limitation, subsequent studies \cite{sugar-pf,sugar-pf-Amrit,infeasibility-quantified-pf-trad} introduced slack variables to quantify power-flow constraint violations and enable convergence in otherwise collapsed systems. Building upon these advances, sparse-optimization techniques \cite{SparseFeas} have been deployed to identify a minimal set of critical locations responsible for system collapse, thereby diagnose dominant sources of key vulnerabilities. Furthermore, the sparse diagnosis framework have been extended to distribution networks, \cite{SparseFeas4DNet}, combined transmission–distribution networks \cite{SparseFeas4TnDNet}, and multi-period growing stress scenario \cite{sparsefeas-mp}. Nevertheless, none of these existing methods incorporate frequency as a system state, a technically challenging yet essential task.

Inspired by these advances and limitations, this work proposes a \textit{frequency-aware sparse optimization framework} to jointly diagnose instability risks and collapses. Specifically, we aim to jointly address the following research questions:
\begin{itemize}[leftmargin=4mm]
    \item \textbf{(Q1) Survivability:} After primary frequency response, can the system reach a healthy steady state (i.e., a feasible power flow solution with an acceptable frequency deviation $\Delta f$)?
\item \textbf{(Q2) Dominant vulnerability:} If not, what key vulnerabilities render the system unstable or lead to its collapse?
\end{itemize}
To answer these questions, we model the frequency state along with governor droop control within power flow constraints, and impose frequency-dependent stability limits defined by user-specified bounds  $\Delta f_{\min}\leq\Delta f\leq\Delta f_{\max}$.  For the first time in the literature, we further enforce sparsity in the solution space to identify the minimal set of compensations required to simultaneously restore power balance and frequency stability. Finally, we adopt I-V based circuit-theoretic formulations and optimization heuristics to efficiently solve the problem at scale.

Findings from this work will advance risk analysis capability during real-time grid operations. The proposed tool is envisioned as a supplement to the traditional contingency analysis within the EMS environments, which are typically performed periodically (e.g., every 30 minutes) using the latest available system state estimates and assumed contingency scenarios. To validate the efficacy of the proposed method, we evaluate the frequency response under significant N-1 generator contingencies across both small and large systems. As shown in Fig. \ref{fig: graph plot, sparsefeas-freq}, only a few localized compensations can stabilize the entire system, restoring frequency within the user-defined safe range. These findings demonstrate that our proposed method mitigates instability risks through localized corrective actions. Our method scales well to large systems, requiring less than 4 minutes to pinpoint sparse solutions for the 1354-bus systems.


\section{Background}
\label{sec:bkg}

\subsection{Circuit-Theoretic Modeling of Power Grid}\label{bkg: ckt formulation}

Recent advances have introduced circuit-theoretic formulations for power-flow and optimization problems, drawing inspiration from classical circuit simulators such as SPICE \cite{sparsePFref15-spice}. Unlike conventional power-based models that rely on $(P,Q,V)$ parameters and/or polar-state variables $(|V|,\delta)$, the circuit-theoretic framework models each component through its current–voltage ($I$–$V$) characteristics in rectangular coordinates using state variables $\bm{v}$ consisting of $(V^{Real},V^{Imag})$. Under this representation, AC network balance $\bm{g}(\cdot)$ is characterized by Kirchhoff’s Current Law (KCL). Notably, this formulation applies consistently across both transmission and distribution networks and has been successfully applied to power-flow analysis \cite{sugar-pf}, state estimation \cite{SUGAR-SE-Li, convexSE-LAV-Li, ckt-GSE,BayesGSE}, and optimal power flow \cite{sparsePFref12-SUGAR-opf}. Building on this foundation, this work adopts circuit-based modeling to incorporate system constraints and design optimization heuristics for large-scale power networks.

\vspace{-2mm}
\subsection{Power Flow Analysis for Collapsed Systems}
Previous studies \cite{sugar-pf,sugar-pf-Amrit} developed a power-flow framework as in \textit{Problem~\ref{eq:infeasibility_dense}}, which converges even for collapsed systems by identifying a minimal set of deficient resources represented by current injections $\bm{n} = [n_i]$, to restore power balance. Here, $\bm{v}=[V_i]$ is a vector of all bus voltages ($V_i=V_i^{\text{Real}}+jV_i^{\text{Imag}}, \forall \text{ bus } i$), and each slack source $n_i$ represents a current injection at bus $i$ within the I-V-based nonlinear AC network constraints $g_i(\cdot)$ (i.e., KCL equations).
\begin{problem}[Dense optimization]\label{eq:infeasibility_dense}
    \begin{equation}
    \min_{\bm{v},\bm{n}}~ \frac{1}{2} ||\bm{n}||_{_2}^2, ~~~\text{ s.t. }~ g_i(\bm{v}) + n_i = 0, ~\forall i\label{prob: sugar pf}
\end{equation}
\end{problem}

\vspace{-4mm}
\subsection{Modeling Frequency in Steady-State Simulation}

Prior work \cite{sugar-pf-freq} extended steady-state power-flow \textit{Problem \ref{prob: sugar pf}} to implicitly model the steady-state frequency deviation $\Delta f$ by incorporating droop control physics. In the resulting formulation, summarized in \textit{Problem~\ref{prob: dense with droop}}, $\Delta P_j$ represents the droop-induced power adjustment at each synchronous generator~$j$, $\bm{\Delta P}=[\Delta P_j]$, and $F_j(\Delta f)$ denotes the governor relationship in droop control for each generator $j$. 

\begin{problem}[Dense Optimization with Droop Resource]\label{prob: dense with droop}
\vspace{-3mm}
\begin{align}
    &\min_{\bm{v}, \bm{\Delta P},\Delta f,\bm{n}} \quad  \tfrac{1}{2}\|\bm{n}\|_{_2}^2 \notag\\
    \text{s.t.} \quad 
    & g_i(\bm{v}, \bm{\Delta P}) + n_i = 0, ~~~\forall i\notag\\
    & \Delta P_j = F_j(\Delta f), ~~~\forall j
\end{align}
\end{problem}

\section{Frequency-Aware Sparse Optimization}
\label{sec:Frequency Aware optimization}
To identify the dominant sources of instabilities, in this section, we integrate frequency modeling and frequency stability bounds within the sparse optimization framework. 

\vspace{-3mm}
\subsection{Frequency-Aware Sparse Optimization}
\label{sec: frequency Aware Sparse Optimization}

The goal of our proposed frequency-aware sparse optimization method, presented as \textit{Problem~\ref{prob: sparse with Freq}}, is to identify the minimal set of locations where compensations not only restore power balance feasibility but also maintain frequency stability following the droop response. When the system is feasible and stable, we will have $||\bm{n}||_{_2}={0}$, and our formulation reduces to  \textit{Problem~\ref{prob: dense with droop}}, which returns the same solution as the traditional power flow and the solution represents the actual system voltage and frequency response to a disturbance.

\begin{problem}[Sparse Optimization with Frequency Bound]\label{prob: sparse with Freq}
\begin{equation}
\begin{aligned}
    &\min_{\bm{v},\bm{\Delta P},\Delta f, \bm{n}} \quad  \tfrac{1}{2}\|\bm{n}\|_{_2}^2  + \sum_i c_i|n_i| \\
    \text{s.t.} \quad 
    & g_i(\bm{v}, \bm{\Delta P)} + n_i = 0, \quad \forall i \\
    & \Delta P_j = F_j(\Delta f), ~~~\forall j \\
    & \Delta f_{\min} \leq \Delta f \leq \Delta f_{\max}
\end{aligned}
\end{equation}
\end{problem}

The frequency stability bound $\Delta f_{\min} \leq \Delta f \leq \Delta f_{\max}$ serves to keep steady-state frequency deviations within a pre-defined safe range (e.g. $\pm 0.3$~Hz). In practice, excessive deviation from the nominal frequency can trigger protective relay actions, leading to generator or load disconnections and, in severe cases, cascading outages or widespread blackouts. As a result, by enforcing this stability criterion as a hard constraint, the sparse solution $\bm{n}$ obtained from \textit{Problem~\ref{prob: sparse with Freq}} identifies the dominant system vulnerabilities responsible for the system's instability; or equivalently, determines the minimal set of compensations that can maintain frequency within the safe range following the droop response.

To incorporate realistic frequency-responsive behavior through a continuous differentiable function, the following relaxed droop-based $\Delta P-\Delta f$ relationship is considered for each generator $j$, as illustrated in Fig. \ref{fig:droop}.
\vspace{-2mm}
\begin{equation}
\label{eqn_droop_bound}
F_j(\Delta f) =
\begin{cases}
\Delta P_{j, \text{min}}, & \Delta f \geq \Delta f_1, \\
a_{1}(\Delta f)^2 + b_{1} \Delta f + c_{1}, & \Delta f_2 \leq \Delta f < \Delta f_1, \\
-k_j \Delta f, & \Delta f_3 \leq \Delta f < \Delta f_2, \\
a_{2}(\Delta f)^2 + b_{2} \Delta f + c_{2}, & \Delta f_4 \leq \Delta f < \Delta f_3, \\
\Delta P_{j, \text{max}}, & \Delta f < \Delta f_4.\notag
\end{cases}
\end{equation}

In the above expression, in the central operating region, generator response is modeled as $\Delta P_j = -k_j \Delta f$, representing a linear  droop control feedback. However, the available droop capability is physically constrained by the generator's minimum and maximum  output limits, i.e., $P_{j,\min}$ and $P_{j,\max}$. Extending a purely linear characteristic to these boundaries introduces non-differentiability at the transition points, which is numerically undesirable. To eliminate these sharp discontinuities, small scale ($\delta =  10^{-8} $) smooth quadratic transition regions are introduced on both sides of the linear segment, resulting in the five-region droop function curve in Fig.~\ref{fig:droop}. 
Quadratic coefficients $a_1,b_1,c_1,a_2, b_2$ and  $c_2$  are obtained by enforcing both power continuity and slope continuity with the adjacent linear segments. This relaxation ensures that the droop curve is continuous and differentiable, eliminating the numerical instabilities in our frequency-aware sparse optimization framework. 

\begin{figure}[htbp]
\vspace{-3.5mm}
    \centering
    \includegraphics[width=0.4\textwidth]{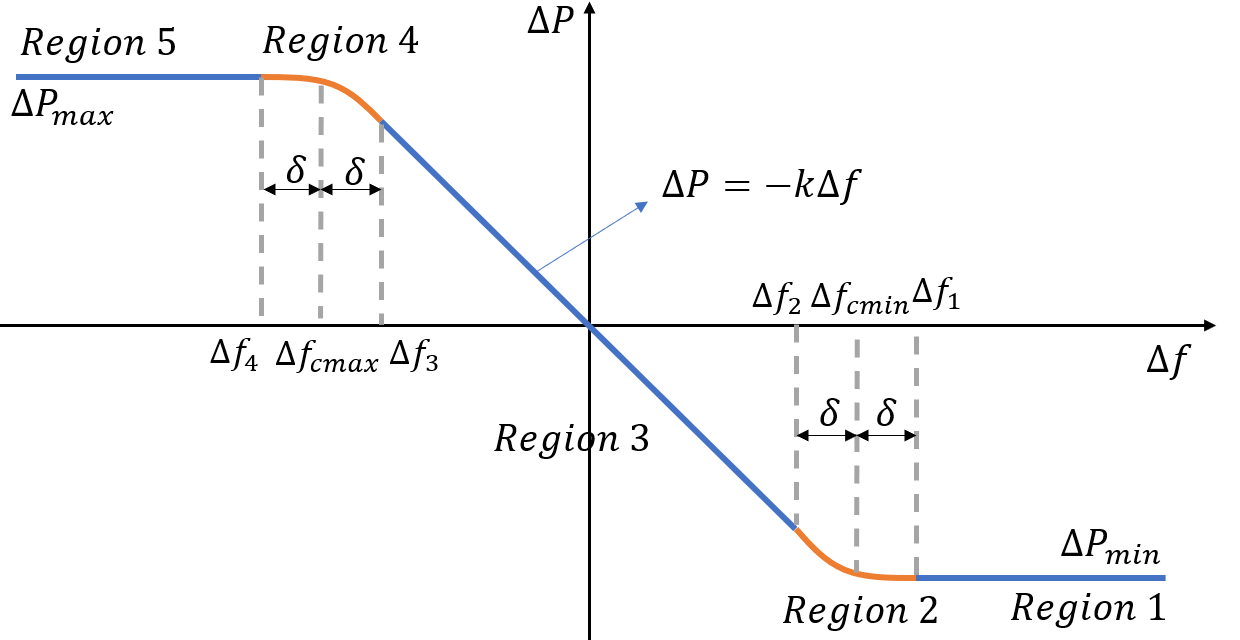} 
    \caption{Droop characteristics of a generator with quadratic relaxation.}
    \label{fig:droop}
    \vspace{-3mm}
\end{figure}






\subsection{Solving \textit{Problem~\ref{prob: sparse with Freq}} at Large Scale}
\label{sec: Solving the problem at large scale}

The proposed \textit{Problem~\ref{prob: sparse with Freq}} is inherently a nonlinear constrained optimization problem, which can be challenging to solve at large scale. To address this, we adopt an interior-point method equipped with proper initialization and circuit-inspired optimization heuristics.
Specifically, we implement a circuit-based interior-point solver, wherein a set of nonlinear perturbed Karush-Kuhn-Tucker (KKT) conditions are formed and solved using the Newton-Raphson method. Circuit-inspired optimization heuristics (including voltage limiting and damping) \cite{sugar-pf-Amrit} are incorporated to enhance convergence stability of interior-point iterations.
Moreover, we aim to find a strictly feasible solution as a starting point for the sparse optimization. This is achieved by solving a dense optimization $\min_{\bm{v}, \bm{\Delta P}, \Delta f, \bm{n}} ~ \tfrac{1}{2}\|\bm{n}\|_{_2}^2$ subject to the same constraints as in \textit{Problem~\ref{prob: sparse with Freq}}.

The need for achieving highly sparse solutions in large systems, while mitigating numerical difficulties, calls for innovative approaches. To this end, we construct a sequence of subproblems  that begins with the dense optimization discussed earlier and gradually increases the sparsity level (by updating the sparsity coefficients $\bm{c}$). This is equivalent to splitting the original problem into a series of subproblems, where each subproblem solves \textit{Problem \ref{prob: sparse with Freq}} with updated sparsity coefficients, using the solution from the previous subproblem as its initial guess. Through this progressive refinement, each subproblem converges to its optimal solution within only a few iterations. At each update step (i.e., when we solve a subproblem with an updated sparsity coefficient), we leverage the efficient sparsity-enforcing mechanism proposed in \cite{SparseFeas}, which adjusts the location-specific sparsity coefficient $c_i$ for each bus $i$. The value of $c_i$ is adaptively toggled between a relatively larger value $c_H = 10$ and a smaller value $c_L = 0.1$ to introduce uneven penalties across locations. This adaptive mechanism promotes highly sparse yet numerically stable solutions. Further details on the selection and theoretical justification of $c_H$ and $c_L$ from an optimization standpoint can be found in our prior work~\cite{SparseFeas}. 

\section{Results}
\label{sec:Results}
\noindent This section presents the experiments conducted to evaluate the efficacy of our proposed method.
We test on standard IEEE cases of transmission systems: Case30, Case300 and Case1354pegase \cite{fliscounakis2013contingency}. For each test case, we evaluate the top $K$ largest N-1 generation outages. A 4\% droop setting is assumed for each active generator, and the frequency stability bounds are set to $-0.3 \text{~Hz}\leq\Delta f\leq 0.3\text{~Hz}$ for Case30, and $-0.06\text{~Hz}\leq\Delta f\leq 0.06\text{~Hz}$ for other cases. 

Notably, although this paper mainly experiments on N-1 generator contingency, the proposed work itself is agnostic to event patterns, and applies generally to any outage or disturbance (e.g., sudden load increase at some location) that can make the system unstable or collapse.

\subsection{Insights into Frequency Instability Risks}

In this study, we have tested and compared two methods: (i) \textbf{Frequency without Bound} which solves baseline \textit{Problem \ref{prob: dense with droop}}; and (ii) \textbf{Frequency with Bound} which solves our Frequency-aware sparse optimization formulated in \textit{Problem \ref{prob: sparse with Freq}}.




Figs. \ref{fig:case30 freq}-\ref{case 1354 freq} illustrate the frequency deviation curves for the systems under N−1 generator contingencies. Plots lying above the lower bounds (red lines) represent \textit{normal/stable} systems that reach feasible (power-balanced) and stable ($\Delta f$ within bounds) steady-states following the droop response. In these scenarios, both the baseline and proposed methods converge to the same solution, requiring no additional compensation sources, with the zero $\bm{n}$ vector indicating system survivability. Whereas, when the system encounters larger generation outages, the plots falling below the lower bound correspond to \textit{unstable} systems, exhibiting frequency deviations that drop well beyond the pre-defined safe range after the droop response. In such cases, our proposed method (Figs. \ref{case30_b}, \ref{case300_b} and \ref{case1354_b}) returns a non-zero sparse $\bm{n}$ vector. The presence of non-zero entries in $\bm{n}$ indicates compensations required at a few identified buses to restore both frequency stability and overall system feasibility. 

\begin{figure}[htbp]
\vspace{-1mm}
    \centering
    \begin{subfigure}[h]{0.4\linewidth}
        \centering
        \includegraphics[width=\linewidth]{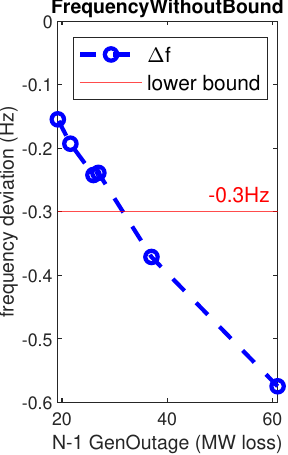}
        \caption{}
        \label{case30_a}
    \end{subfigure}
    \begin{subfigure}[h]{0.4\linewidth}
        \centering
        \includegraphics[width=\linewidth]{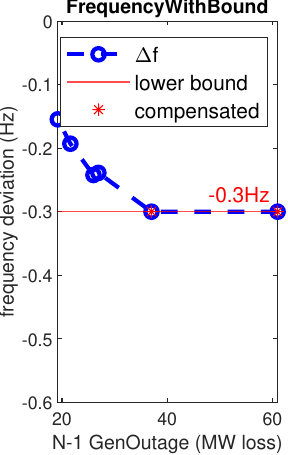}
        \caption{}
        \label{case30_b}
    \end{subfigure}
     \hfill
    \caption{Case30 frequency curve, N-1 generator contingency.}
    \label{fig:case30 freq}
    \vspace{-1mm}
\end{figure}
\begin{figure}[htbp]
\vspace{-2mm}
    \centering
    
    \begin{subfigure}[h]{0.4\linewidth}
        \centering
        \includegraphics[width=\linewidth]{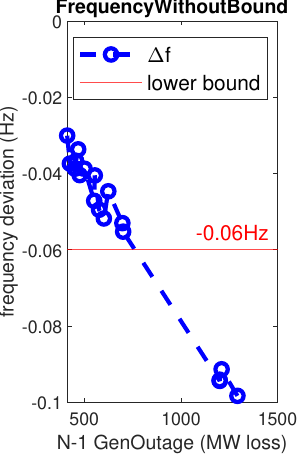}
        \caption{}
        \label{case300_a}
    \end{subfigure}
    \begin{subfigure}[h]{0.4\linewidth}
        \centering
        \includegraphics[width=\linewidth]{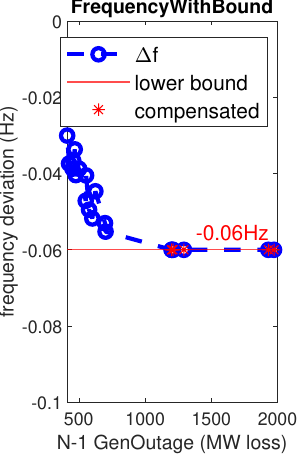}
        \caption{}
        \label{case300_b}
        \vspace{-2mm}
    \end{subfigure}
     \hfill
    \caption{Case300 frequency curve under N-1 generator contingencies. Cases with $>$ 1500MW loss are not shown in (a) because the system completely collapses and the frequency deviation becomes non-informative. In (b), these cases are recovered to a user-defined \textit{healthy} state through our proposed method.}
    \label{case 300 freq}
\end{figure}
\begin{figure}[htbp]
\vspace{-2mm}
    \centering
    \begin{subfigure}[h]{0.4\linewidth}
        \centering
        \includegraphics[width=\linewidth]{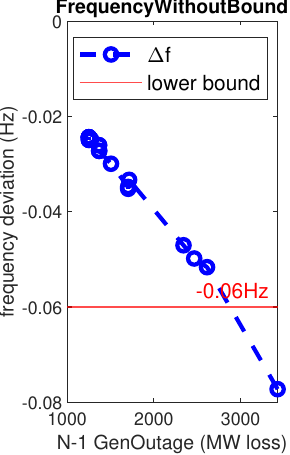}
        \caption{}
        \label{case1354_a}
    \end{subfigure}
    \begin{subfigure}[h]{0.4\linewidth}
        \centering
        \includegraphics[width=\linewidth]{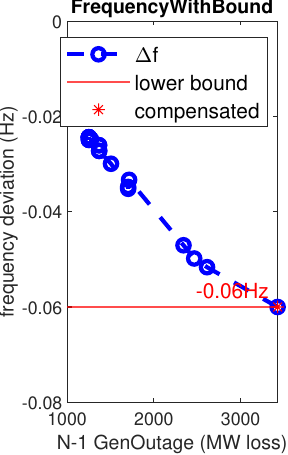}
        \caption{}
        \label{case1354_b}
    \end{subfigure}
     \hfill
    \caption{Case1354pegase frequency curve under N-1 generator contingency.}
    \label{case 1354 freq}
    \vspace{-3.5mm}
\end{figure}








\vspace{-3mm}
\subsection{Speed and Scalability}
We evaluate the runtime performance across standard systems of varying sizes, ranging from 30-bus to 1354-bus networks. As shown in Fig. \ref{fig:scalability}(a) tests on normal conditions where system remains feasible and stable after generator outage.  
Whereas, Fig. \ref{fig:scalability}(b) tests unstable conditions where system reaches a feasible steady state after generator outage but the frequency deviation $\Delta f$ violates the pre-defined bound. 
\begin{figure}[htbp]
\vspace{-2mm}
    \centering
    \includegraphics[width=\linewidth]{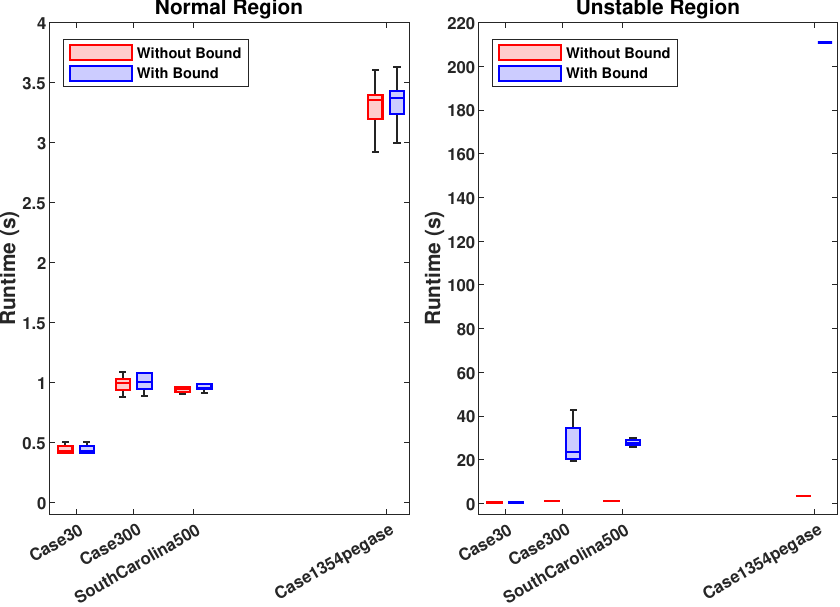} 
    \caption{  Scalability. Left plot: under normal system conditions, runtime of baseline and proposed methods are comparable because the system does not require any compensation and both methods converge quickly (to $\bm{n}=\bm{0})$. Right plot: for the unstable conditions, the
proposed method will require longer time as
it involves extra outer-loops to pinpoint the sparse vulnerability
locations. }
    \label{fig:scalability}
    \vspace{-2mm}
\end{figure}



\section{Conclusion}
\label{sec:Conclusion}
This paper presented a frequency-aware optimization framework for evaluating the survivability of power systems under generator tripping disturbances. By analyzing steady-state frequency deviations after droop control, the proposed method identifies a minimal set of locations along with necessary compensations that reveal the system vulnerabilities responsible for instability or collapse. For example, in a 1354-bus system under an N-1 generator contingency (3424.8 MW loss), the system’s steady-state frequency deviation after droop response drops by nearly -0.08 Hz, whereas the proposed method effectively finds four dominant vulnerability locations  (Bus IDs: 701, 734, 763 and 952) and determines corresponding compensations as slack sources to ensure both power balance and frequency stability (within -0.06 Hz limit). The method also scales well with system size, requiring less than 4 minutes to pinpoint sparse solutions for the 1354-bus systems on a laptop computer. To further enhance practicality, the proposed method is  extensible to incorporate advanced frequency regulation technologies (e.g., virtual synchronous generators), automatic voltage regulators (e.g., FACTS devices), and can further extend to inaccurate, uncertain, or partially known system models / states arising from real-world situational awareness limitations, which we reserve for future research.



\bibliographystyle{IEEEtran}
\bibliography{main}

\end{document}